# New Optical Link Technologies for HEP Experiments

P. Delurgio, W. Fernando, B. Salvachua, R. Stanek, D. Underwood
*High Energy Physics, Argonne National Lab, Argonne IL 60439, USA*
D.Lopez
*Center for Nanoscale Materials, Argonne National Lab, Argonne IL 60439, USA*

As a concern with the reliability and mass of current optical links in LHC experiments, we are investigating CW lasers and light modulators as an alternative to VCSELs. In addition we are developing data links in air, utilizing steering by MEMS mirrors and optical feedback paths for the control loop. Laser, modulator, and lens systems used are described, as well as two different electronic systems for a free space steering feedback loop. Our prototype system currently operates at 1.25 Gb/s, but could be upgraded. This link works over distances of order meters. Such links might enable one to move communication lasers (e.g. VCSELs) and optical fibers out of tracking detectors, for reasons such as reliability and power consumption. Some applications for free space data links, such as local triggering and data readout and trigger-clock distribution and links for much longer distances are also discussed.

## 1. Introduction

The detector elements of future High Energy Physics (HEP) experiments will collect large amounts of data and there is a need to find ways to get the data out of the detectors efficiently and reliably while at the same time reducing the mass of the communication electronics. Optical communication is the only option for transferring data due to the required bandwidth. Currently modulated lasers are commonly used in HEP experiments, but in the future Electro Optical Modulators (EOM) will be used because of higher bandwidth (no chirp), low power, and reliability. Beams in air can be important for eliminating fiber routing. Also, the trigger latency can be reduced because of the velocity factor for beams in air relative to beams in fiber.

## 2. Electro Optical Modulators (EOM)

A light beam can be modulated in phase, frequency, amplitude, or polarization. Amplitude modulation is the most popular method in optical communication. EOM uses a material with an electro-optic effect to modulate a beam of light with an electrical signal. There are two basic kinds of EOMs of interest, Electro Absorption Modulators (EAM), and Mach-Zhender Interferometer based modulators (MZI). Lithium Niobate ($LiNbO_3$) with various doping has been used in MZI based EOMs for a long time. In $LiNbO_3$ crystals the refractive index changes as a function of the strength of the local electric field. If it is exposed to an electric field, the refractive index will increase and light will travel slowly through it. Therefore, the phase of the light going through the crystal is controlled by the local electric field in the crystal. The phase modulated beam can be converted to amplitude modulated beam by combining two beams; one modulated in phase, and the other with the same amplitude but without phase modulation.

$LiNbO_3$ has been tested for radiation hardness by several HEP groups [1]. The only disadvantage for LiNbO3 is size (few cm long). There are commercial modulators of small size, but some are polymer (not radiation hard) and some are too expensive at the present time. We may have found two vendors for small modulators who will work with us on ones which can be wire-bonded and have single-mode fiber connections (shown in Figure 1(a)). Radiation hardness tests need to be performed on these to determine suitability for HEP detectors.

Recently, with the discovery of silicon waveguides [2], many EOMs have been developed based on semiconductors. These EOMs are extremely small in size, ultra low in power consumption and can be integrated into CMOS. Luxtera/Molex monolithically fabricated a 0.13 μm CMOS Silicon On Insulator (SOI) integrated 40 Gb/s single chip MZI modulator based transceiver [3]. This is a 4 channel transceiver with the drivers, modulators and PIN diodes; all are integrated in to a single CMOS. A functional block diagram and a photo of a CMOS die are shown in Figure 2. The commercial version of this chip has a CW laser integrated at the packaging, but the laser could be removed and replaced with an optical fiber for HEP applications. InP based single channel 40Gb/s MZI modulators from Heinrich-Hertz-Institute (HHI) [4] (Figure 1(b)) and GaAs based $2 \times 20$Gb/s MZI modulators from u²t Photonics [5] (Figure 1(c)) are also semiconductor modulators commercially available with very small footprints.

More light modulators are under development at Intel [6], IBM [7] and MIT [8]. The Intel design is a 40Gb/s MZI Si based modulator which uses wavelength-division multiplexing (WDM) to reach 50Gb/s in a single fiber and all the components of the transceiver is integrated into a single die. Since the laser is integrated in this die, use of this chip in our setup is not possible without major modifications. The MIT prototype uses Si stresses with Ge which has an absorption peak which is used to modulate the light. The radiation hardness of GeSi in general has been studied [9, 10] and shown to be adequate for use in future HEP experiments. This device is fabricated with 180 nm CMOS technology with a very small footprint (30 μm$^2$). The IBM prototype is an extremely small MZI based EOM. A graphene based EAM was designed by the National Science Foundation (NSF) and Nanoscale Science and Engineering Center at the University of California-Berkeley with a footprint of 25 μm$^2$ and has a possibility of running up to 500 Gb/s independent from temperature or the frequency of the laser (in the range of 1.3 μm – 1.6 μm) [11].

Many commercial systems which work faster than 10 Gb/s already use EOMs and CW lasers. EOMs enable one to get the lasers out of tracking volume.

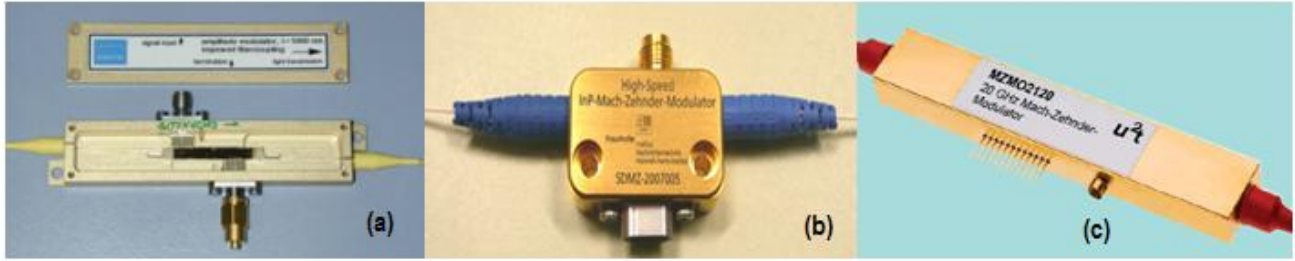

Figure 1: Different EOM available for testing (a) LiNbO$_3$ base EOM from Jeoptic (b) InP base EOM from HHI (c) GaAs based EOM from u$^2$t

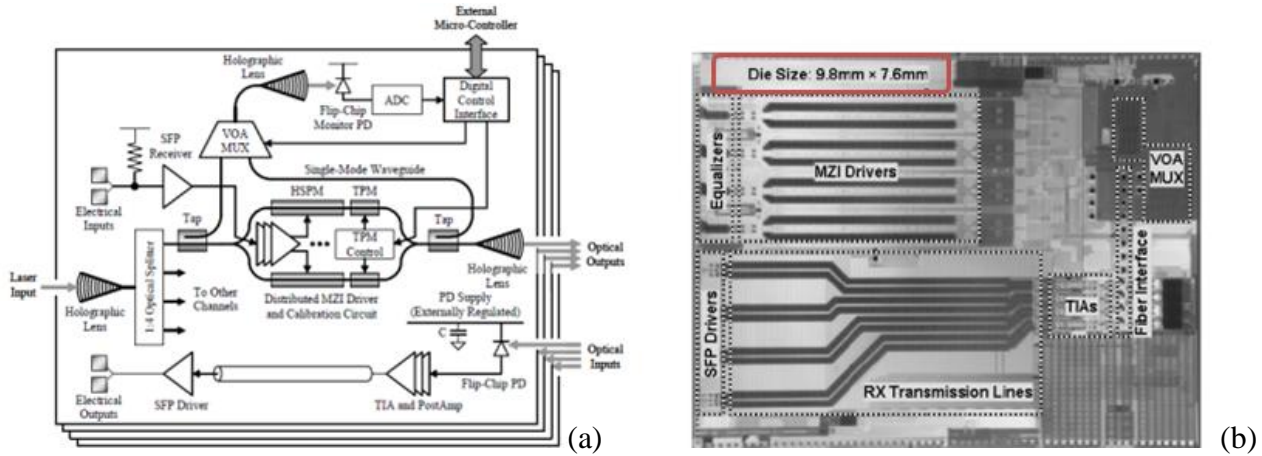

Figure 2: Luxtera/Molex 0.13 μm CMOS SOI integrated 40 Gb/s single chip MZI modulator based transceiver [3]. (a) Block diagram (b) Photo of the CMOS layout

## 3. The Comparison of VCSELs and EOMs

Many current HEP experiments use Vertical Cavity Surface Emitting Lasers (VCSELs) in both sides of the optical communication links to generate modulated light including ATLAS [12] and CMS [13] pixel detectors. Since each VCSEL must be operated at a critical operating point, a radiation hard Application Specific Integrated Circuit (ASIC) has been developed for use with VCSELs. A conventional laser transceiver uses about 300 mW of power at 1 Gb/s. A VCSEL with its driver can be of ~100 mW. An MIT prototype modulator used 50 μW of electric power. This is a huge improvement in power consumption (and heat generation). The existing VCSEL and driver ASIC is ~ cm long, but the prototype modulators from both MIT and IBM are about $10^{-2}$ cm long (100 μm), a volume ratio of ~$10^{-4}$ and a mass ratio of similar order.

VCSELs used in these links were chosen after studying the radiation hardness of the commercial VCSELs [14] which may or may not satisfy the radiation hardness required for the future HEP experiments. Since there is evidence of high failure rates of VCSELs [15], which may or may not be associated with particular manufacturers or particular designs, it is becoming clear that it is important to move the lasers used for data transmission out of areas inaccessible after closing of the detector to run the experiment.

## 4. Data Link Conceptual Design

In this design all the laser sources are moved outside the detector and EOMs are used to transfer the data into light beams (see Figure 3). This not only reduces the current required inside the detector but also removes the necessity to develop complex radiation hard current driver ASICs. CW laser sources are used instead of VCSELs. Similar to the Luxtera/Molex design, the complete transceiver could be built on a single die and finally the on-detector transceiver could be integrated to the last stage of the sensor/module control chip. The current optical communication scheme has many different components (VCSELs, current drivers, PIN diodes, trans-impedance amplifiers etc.) wire bonded. This design eliminates the reliability issues associated with the loose wire bonds since all the components are integrated. The fibers between the modulators and photo diodes can be replaced with beams in air if it is required.

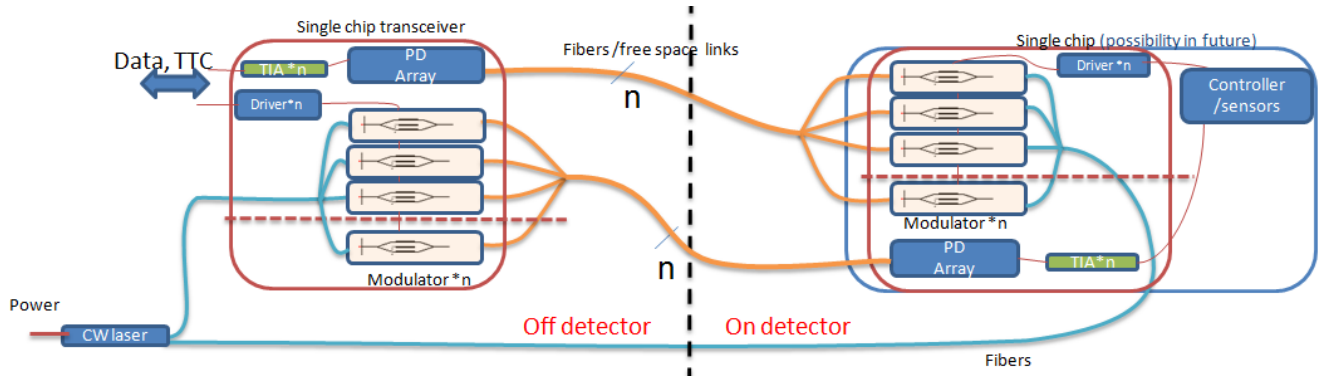

Figure 3: Illustration of conceptual design

## 5. Future Plan with EOMs

A study of the radiation hardness of commercially available modulators is planned to investigate the possibility of using these commercial modulators for near future HEP experiments. A selection of different types (LiNbO3, Luxtera/Molex SOI, and InP) of EOMs are planned to be tested for Total Ionization Dose (TID) with elections at Argonne. Subsequently, testing for lattice displacement and Single Event Effects (SEE) using 200 MeV protons up to $10^{15}$ p/cm$^2$ should occur by the end of 2011. For the longer term a collaboration will be formed between ANL, Fermi National Laboratory and a commercial company to develop a reliable fast radiation hard EOM for HEP applications.

## 6. Free Space Data Links

Instead of fibers the light beams could be steered in air into the modulators by using MEMS mirrors and special light couplers. Since fibers are radiation sensitive, particularly in areas where there is very high rate of exposure, the free space beam suffers no consequence over the lifetime of the experiment. Also free space beams reduce the complexity of fiber routing while also reducing the latency due to velocity factors and delay drift due to thermal effects compared to fibers. There are also areas where fiber connectors are too large and too massive.

### 6.1. MEMS Mirrors

A demonstration steering of a free space beam carrying data has been done using MEMS mirrors at Argonne. If there are relative movements of the sender and receiver, for example vibrations caused by pumps, some sort of method to compensate for the movement must be incorporated into the steering. A steering feedback loop utilizes a position sensitive detector which senses the reflected beam from the mirror at the receiving side. This is illustrated in Figure 4. Laser light is reflected off of a MEMS mirror and onto a partially silvered lens [10]. The lens is chosen to produce a large enough reflection at four simple Si photo detectors rigidly coupled around the MEMS mirror. The change in response is then used as a feedback to move the beam back to the receiver accurately. An analog control loop was used to prove the feedback. Our prototype setup uses a commercial MEMS mirror from Mirrorcle Tech [16].

In order to simulate the relative vibrations, the reflecting lens is mounted on a small rigid structure which can drive at variable frequencies and amplitudes in a single axis. An analog filter was introduced to reduce the vibrations associated with a mechanical resonance of the MEMS mirror. The design and behavior of the analog filter is further explained in [10]. This setup is capable of constraining beam motion relative to target to about 5 μm, when the reflecting lens is moving 700 μm at 5 Hz [10].

There are several issues with this scheme in getting the light beam captured after it passes through the half silvered lens. The first issue is that both the size of the reflection on the silicon detectors and the size of the beam at the receiving lens are dependent on the size of the beam at the reflecting lens. The second issue is that the reflected light pattern on the silicon detectors is not proportional to the movement of the reflecting lens if the movement is large (~mm).

In order to solve the issues with the direct feedback the setup was changed. Instead of a simple feedback a lookup table on an FPGA was used to convert the reflected signal on the silicon detectors to the actual position of the receiver. Here it was essential to introduce a separate laser to do the alignment since the lookup table was dependent on the receiver location (different lookup table for each location). With the separation of the laser, the reflecting lens was moved to the side of the receiving lens. An illustration of the new setup is shown in Figure 5 and photos of the actual setup are shown in Figure 6. The lookup table is filled initially by mapping the reflection patterns to the actual location of the receiving system.

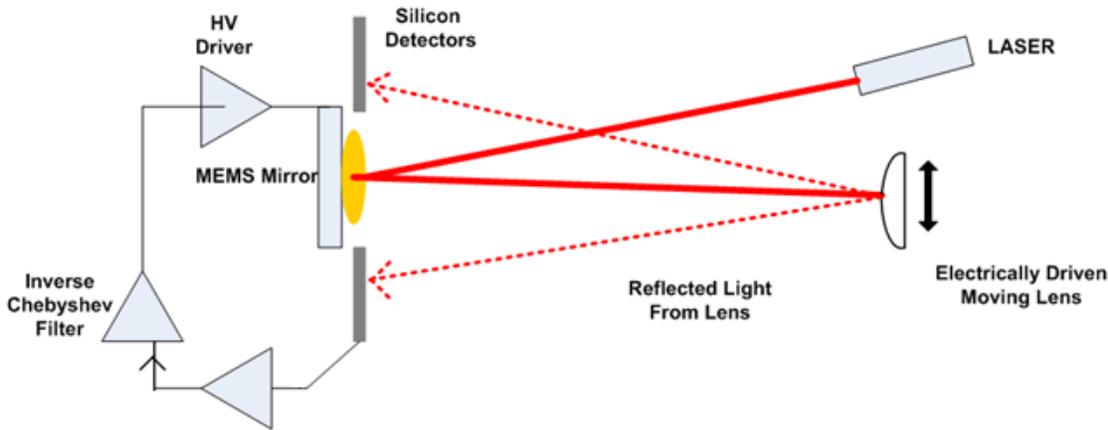

Figure 4: Concept of direct feedback to establish and maintain stable alignment

## 6.2. The Lens System

With the free space data link over a distance of ~1 m, we investigated the ability of different types of lenses to form the beam and capture the beam at various angles and positions. To build a complete link without major optical losses, it is essential to optimize both beam launch and capture. Aspheric lenses were found to produce narrow beams but have very narrow acceptance angles (< 2 mr). GRIN lenses were found to have wide acceptance angles (>20 mr) if a multi mode fiber is attached. With these findings our setup used an aspheric lens to launch the beam and a GRIN lens to capture the beam.

## 7. Prototype Setup

A prototype was built to test the lens system, MEMS mirrors and feedback all together, as illustrated in Figure 5. A pseudo random digital electrical signal is generated from an FPGA board at 1.25 Gb/s and that signal was used to modulate a 1550 nm CW laser beam with an EOM. For this setup a $LiNbO_3$ MZI type EOM was used. The modulated light is then launched (in air) using an aspheric lens. The beam is reflected off of a MEMS mirror in the direction of the GRIN lens in the receiving setup. The light captured by the GRIN lens is then fed to a 1550 nm Small Form-factor Pluggable (SFP) transceiver to convert the optical signal back to electrical, which feedbacks to the FPGA board to compare with the original signal to check the Bit Error Ratio (BER).

An 850 nm laser was used for the alignment link due to the fact that simple inexpensive Si detectors could be used to detect the reflected laser light. The alignment laser is reflected from a silvered lens which is rigidly coupled to the receiving system of the data link. Any vibration in the receiving system also vibrates this reflecting lens and that changes the reflected pattern at the Si detectors. The Si detectors are connected to an amplifier and digitizer. The digitized signal is then fed to an FPGA, which was used to lookup the MEMS mirror voltages required to steer the beam directly in to the receiving GRIN lens. The signals are converted to analog and passed through inverse Chebyshev low pass analog filters. These filters are required to reduce any 1 and 3 kHz signal, which are resonant frequencies of the MirrorcleTech MEMS mirror.

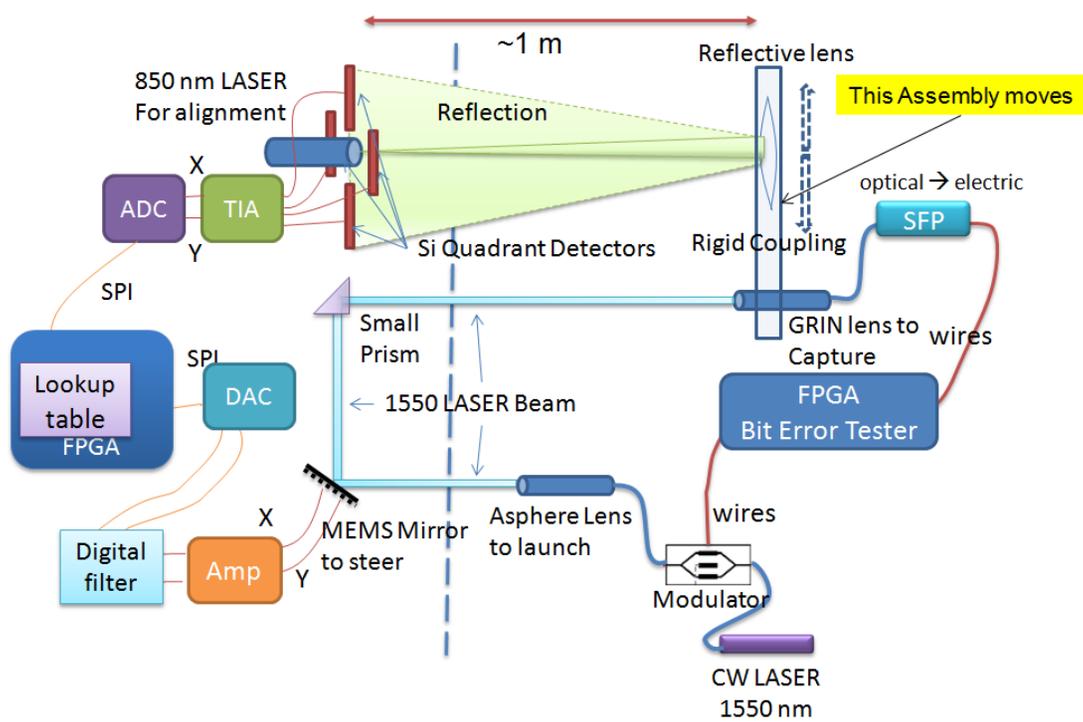

Figure 5: An illustration of demonstration setup

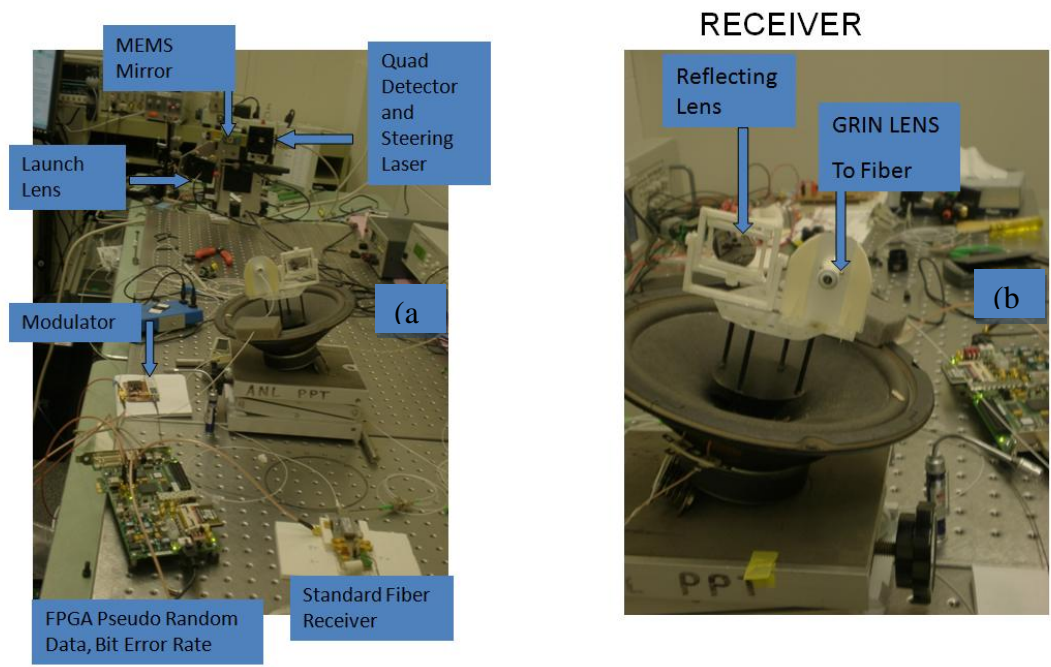

Figure 6: The prototypes setup with all the components (a) Full setup. (b) A closer look at the receiving system and its mounting

## 8. Performance of the System

The static performance (how far the receiver could move) of the full system was measured by moving the receiver very slowly in both the x and y directions. The Full Width Half Maximum (FWHM) of the received optical power was measured and the result is show in Figure 7. The FWHM of optical power was found to be at ~12 mm displacement. The BER at ±12 mm is measured as $<10^{-15}$.

To measure the dynamic performance, the receiving system was mounted on a speaker (Figure 6(a)) at some random angle. The frequency of the speaker was changed keeping the amplitude of the speaker at its maximum at any given frequency. Here the amplitude was not constant and hence we recorded the amplitude and the frequency and conducted a fast BER test ($<10^{-11}$). We chose a few extreme points and conducted a long BER test ($<10^{-15}$). The plot of tested BER at different frequencies and amplitudes is shown in Figure 8. The shape of the figure only reflects the dynamics of woofer response. The system was capable of handling the full spectrum of frequencies and amplitudes that the speaker could produce.

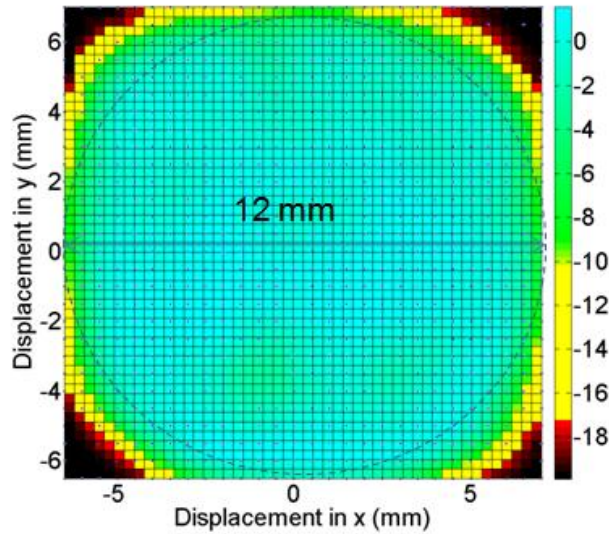

Figure 7: Captured optical power (dBm) with displacement on x and y axis

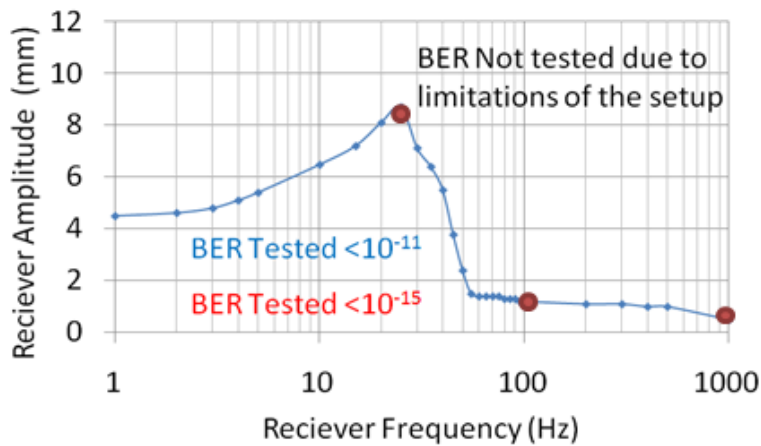

Figure 8: Tested working range of the moving receiver. In this plot the BER was only tested for the points below the blue line; limitations in the setup restricted movement of the receiver below the blue line.

To understand the performance, we measured the instantaneous captured light power with the same amplitudes and frequencies that we tested the system for BER. The optical power limit required for the SFP receiver is ~-18 dBm, while the lowest optical power captured by the system is greater than -5 dBm as shown in the Figure 9. The measurement reconfirms that the system is capable of handing the entire amplitude/frequency spectrum. The tested working range of the system (~ cm of slow vibration and ~mm fast vibration) can cover the possible vibrations associated with the HEP experiments.

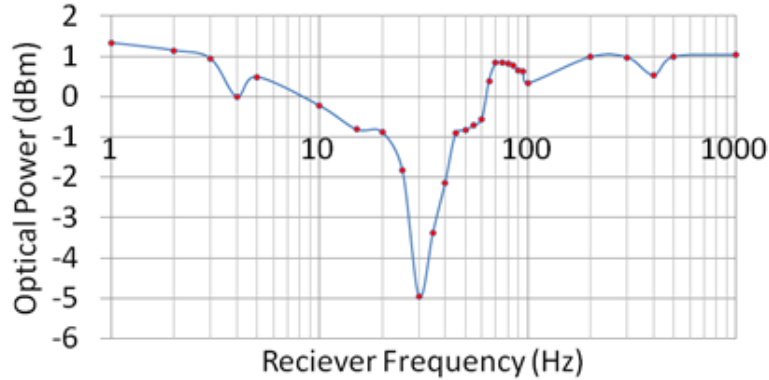

Figure 9: Received optical power at highest amplitude in different frequencies.

## 9. Triggering

On-board triggers would reduce the amount of data to be transmitted out of the tracking detectors. This is essential for LHC upgrades. We show a concept for interlayer communication [10] which would function for the spacing of tracking layers from roughly one cm to several cm as shown in Figure 10. This would allow some momentum cut on track stubs. A major improvement beyond even the conventional form of optical links could be made by using optical modulators so that the lasers are not in the tracking volume.

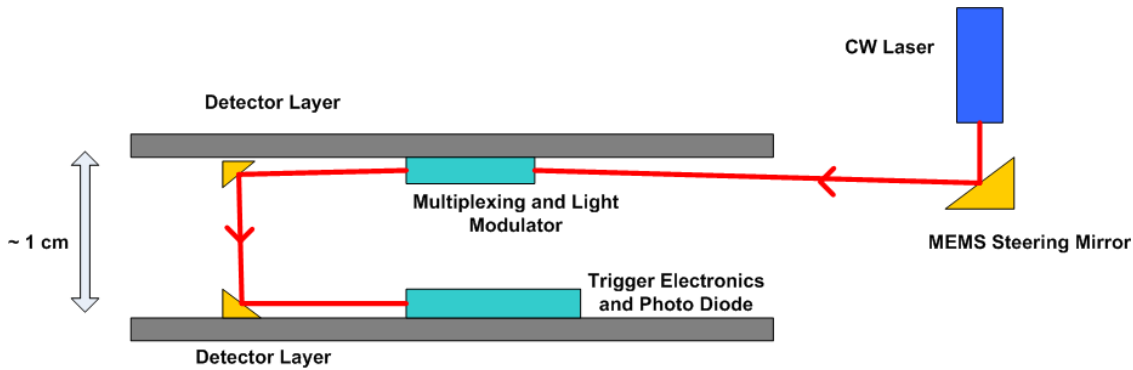

Figure 10: Concept of communication between ID layers for trigger decisions

## 10. For Long Range Communication

Due to diffraction, there is an optimum diameter for a beam for a given distance in order to order to not be in the $1/r^2$ loss region. The Rayleigh distance relates waist size and divergence, and it depends on wavelength. If we start with a diameter too small for the distance of interest, the beam will diverge, and will become $1/r^2$ at the receiver, and we will have large. This is typical of space, satellite, etc. applications. If we start with an optimum diameter, the waist can be near the receiver, and we can capture almost all the light and focus it to a small spot. Examples are ~ 1 mm for 1 m and ~ 50 mm for 1 km.

We have developed a telescope, where the 1.25 Gb/s beam is expanded with a negative lens and uses a mirror to form a beam with a 2 inch diameter, successfully sending it ~ 80 m and capturing it back to an avalanche photo diode. The transmission was verified with eye diagrams with a good opened eye.

## 11. Conclusions

We have made a number of advances in free space optical data transmission at ANL-HEP. Among these are steering using reflections from the receiver system without wires. A major improvement was made by separating data link and alignment link. Ways to form beams and receive beams that reduce critical alignments were developed, reducing time and money for setup. A free space system is operating at 1.25 Gb/s over 1550 nm, using a modulator to compose data, and an FPGA to check for errors, a $<10^{-15}$ error rate, with target moving about 1.2 cm x 1.2 cm at 1.2 m. We have control of a MEMS mirror which has high Q resonance (using both analog and digital filter).

Some future directions are to develop at least a 5 Gb/s free space link (with digital feedback), incorporate more than one MEMS mirror in our setup, and pursue a more robust long distance optical link using steering feedback and micro-motors. MEMS mirrors supplied by Argonne CNM will be evaluated, and radiation tests performed on commercial modulators.

In addition, we have submitted a proposal to apply optical readout to an actual detector in the Fermilab test beam using the Argonne DHCAL [17], which would be an ideal test-bed with 400K channels.

During the course of our studies it has become evident that the use of light modulators would eliminate many of the reliability problems associated with VCSELs, whether with free space transmission or with fibers. We are planning to study the radiation hardness of some commercial EOMs to use in near future HEP experiments. A collaboration to study and design a radiation hard EOM for HEP experiments is in the process of formation.

## Acknowledgements

The authors would like to thank the HEP Division electronics group for assisting in the filter development and frequency measurements. This work was supported in part by the U.S. Department of Energy, Division of High Energy Physics, under Contract KA-15-03-02.